\documentclass[aps,twocolumn,secnumarabic,balancelastpage,amsmath,amssymb,nofootinbib,floatfix]{revtex4-1}

\usepackage{algorithm}
\usepackage{algorithmic}

\usepackage{graphicx}      
\usepackage{float} 
\usepackage{subcaption}
\usepackage{float}
\usepackage{bm}            
\usepackage[colorlinks=true]{hyperref}  

\usepackage{cleveref}

\begin{document}
\title{Unsupervised Learning Approach to Anomaly Detection in Gravitational Wave Data}
\author{Ammar Fayad}
\email{afayad@mit.edu}
\thanks{Preprint: work-in-progress}
\date{October 30, 2024}
\affiliation{MIT Department of Physics}

\begin{abstract}
Gravitational waves (GW), predicted by Einstein’s General Theory of Relativity, provide a powerful probe of astrophysical phenomena and fundamental physics. In this work, we propose an unsupervised anomaly detection method using variational autoencoders (VAEs) to analyze GW time-series data. By training on noise-only data, the VAE accurately reconstructs noise inputs while failing to reconstruct anomalies, such as GW signals, which results in measurable spikes in the reconstruction error. The method was applied to data from the LIGO H1 and L1 detectors. Evaluation on testing datasets containing both noise and GW events demonstrated reliable detection, achieving an area under the ROC curve (AUC) of 0.89. This study introduces VAEs as a robust, unsupervised approach for identifying anomalies in GW data, which offers a scalable framework for detecting known and potentially new phenomena in physics.
\end{abstract}

\maketitle

\section{Introduction}

Gravitational waves (GW) were first predicted by Einstein following his formulation of a general theory of relativity (the theory that describes space and time characteristics given the distribution of energy (and equivalently matter) and momenta). The existence of GW was experimentally confirmed a century later in 2015 by the Laser Interferometer Gravitational-Wave Observatory (LIGO). The first detection of a GW was the event GW150914 which was a result of the merger of two binary black holes. Following this discovery, the LIGO collaboration was awarded the Nobel Prize in Physics in 2017.

The discovery was remarkable for many reasons. Notably, gravitational waves may propagate unimpeded across the cosmos, unlike their electromagnetic counterpart which can be absorbed and scattered by matter. This property allows us to gain information about cosmological systems. It also enables physicists to test and expand the limits of our best understanding of gravity.


In this work, we propose a novel unsupervised learning approach to detect anomalies in gravitational wave time-series data. We propose to use autoencoders (AE) which are deep neural networks trained to minimize a measure\footnote{See \cref{ML} and \ref{method} for example of distances.} of distance between its input and output. The method used in this paper relies on the central idea that if we train an AE on a stream of homogenous data (the noise of the detectors), AE will be able to reconstruct the inputs almost perfectly should they be of the same nature as the training data. However once an anomaly (GW event) occurs in the data, the AE will fail to reconstruct it correctly since it never was included in the training. This ``fail-to-reconstruct" result can be quantitatively assessed to decide whether an anomaly is present \cite{zhou2017anomaly, ibrahim2022hierarchical, finke2021autoencoders}. 


\section{Theory}
This section provides a brief overview of General Relativity and Machine Learning in the context relevant to our experiment.
\subsection{General Relativity}
The standard formulation of General Relativity is encoded in the Einstein's Field Equation (EFE):
$$G_{\mu\nu} = 8\pi G T_{\mu\nu}$$
where $G_{\mu\nu}$ is called the Einstein Tensor and it describes the spacetime geometry (or curvature as it is comprised of second derivative terms of the Riemannian metric $g_{\mu\nu}$). On the other hand, $T_{\mu\nu}$ is the stress-energy tensor that describes the distribution (i.e. density and flux) of energy and momenta in space and time. In the weak-field limit (i.e. the spacetime is almost flat), EFE become a wave equation  
$$\partial_\alpha\partial^\alpha h_{{\mu\nu}} = -16\pi G T_{{\mu\nu}}$$
where $ h_{{\mu\nu}}$ is the strain tensor and describes the small perturbation to the flatness of the spacetime geometry. The above is a wave equation due to the d'Alembertian acting on $h$. The solutions of this equation are known as the gravitational waves with strain $h$.

\subsection{The LIGO detector}\label{ligo-sec}

\begin{figure}
    \centering
    \includegraphics[width=1\linewidth]{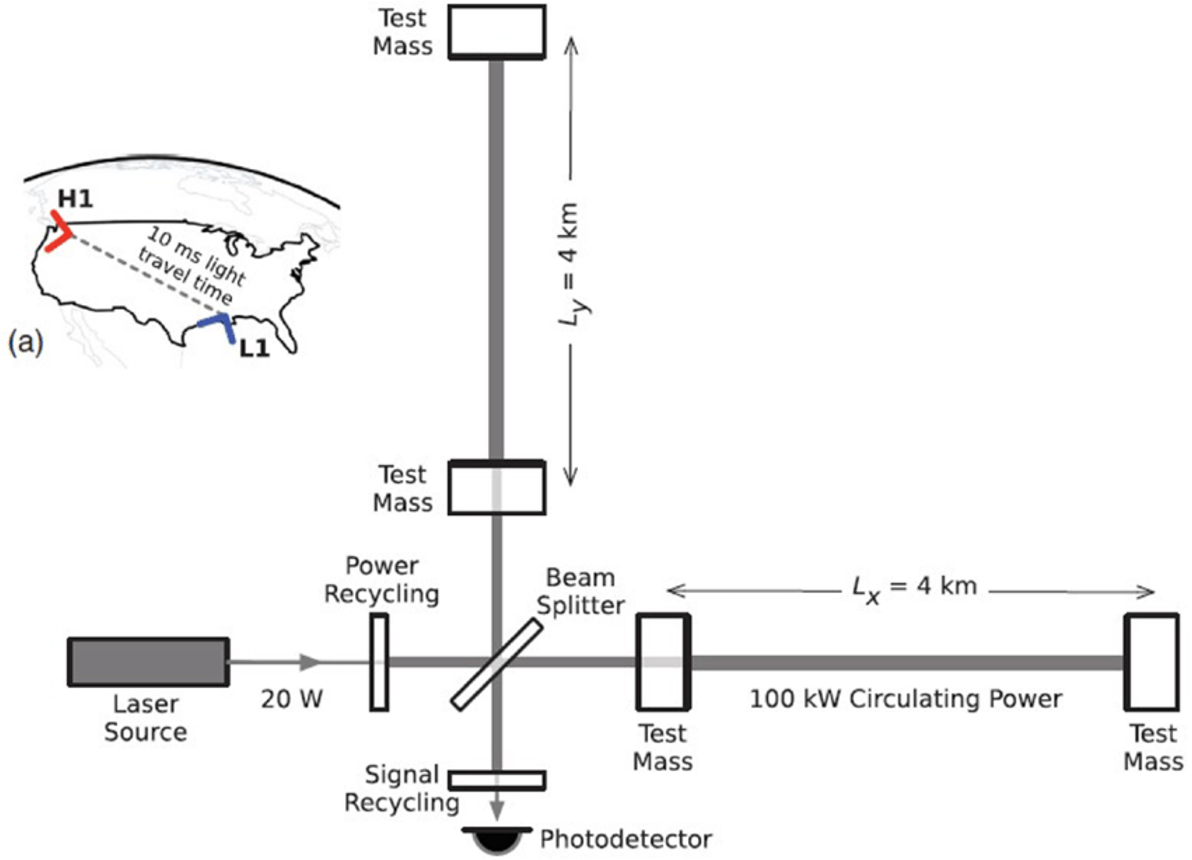}
    \caption{Simplified Diagram of the LIGO detector \citep{abbott2016observation}. The detection mechanism is explained in \cref{ligo-sec}.  Subfigure (a) shows the locations of the L1, and H1 detectors used for our analysis. }
    \label{ligo}
\end{figure}

\begin{figure}
    \centering
    \includegraphics[width=\linewidth]{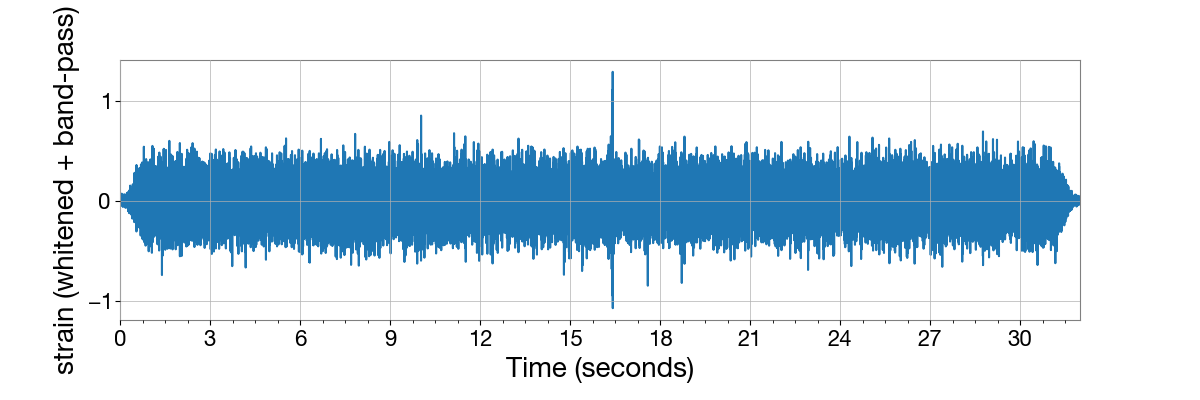}
    \caption{Strain data of event GW150914. The GW signal occurs at time = $16.5$ s. The data is whitened and band-passed in the range $[20,400]$.}
    \label{strain}
\end{figure}

The LIGO detectors (\cref{ligo}) are designed to measure the strain, that is caused by passing gravitational waves (GWs) \citep{abbott2016observation}. Using laser interferometry, these detectors measure relative changes in the lengths of two perpendicular arms with great precision, i.e. strains as small as $10^{-21}$ meters. The detection mechanism operates as follows: a laser beam is split and directed along the two arms, where it reflects off mirrors and recombines at the beam splitter. In the unperturbed conditions, the equal lengths of the arms result in no net interference at the photodetector. However, when a gravitational wave, with a plus polarization for example, passes through, it induces an alternating contraction and dilation of the arms. This creates a phase difference in the laser beams when they recombine, leading to an interference pattern with an intensity proportional to the strain $h$ of the GW. For this study, we analyze strain time-series data, which is publicly available from the Gravitational Wave Open Science Center (\url{https://gwosc.org/}). Refer to \cref{strain} that shows the processed strain timeseries data of the event GW150914. Further details on the processing can be found in \cref{training}.

\subsection{Machine Learning}\label{ML}
We train a deep autoencoder (AE) neural network to detect anomalies. In the simplest form, Neural Networks are statistical learning architectures designed to approximate relationships within data; the main distinction between NNs and simple models such as linear regression is the model complexity where NNs process data sequentially with layers, each consisting of a neuron (or unit) that applies an appropriate transformation on its input. The basic AE model trains on data of the form $\{(x_i,x_i)\}_{i=1}^n$ and learns a function
$f(x_i)$ such that $f(x_i)\approx x_i$. In statistics terminology, the AE learns parameters $\theta$ such that $\mathbb{E}[||f_\theta(X)-X||^2]$ is minimized. AE's main assumption is that the input data can be modeled using a low-dimensional latent variable $Z$. Following this assumption, AE's architecture is as follows: Given input data $X$ of dimension $d_X$, find the latent representation $Z=\text{Encoder}(X)$ which has a lower dimension $d_Z$. Then, output $\hat{X}=\text{Decoder}(Z)$ which has the same dimension as the input $d_X$. In summary. AE = Encoder + Decoder, which are two separate neural networks.


\section{Proposed Methodology}\label{method}

\begin{figure*}
    \centering
    \includegraphics[width=0.72\linewidth]{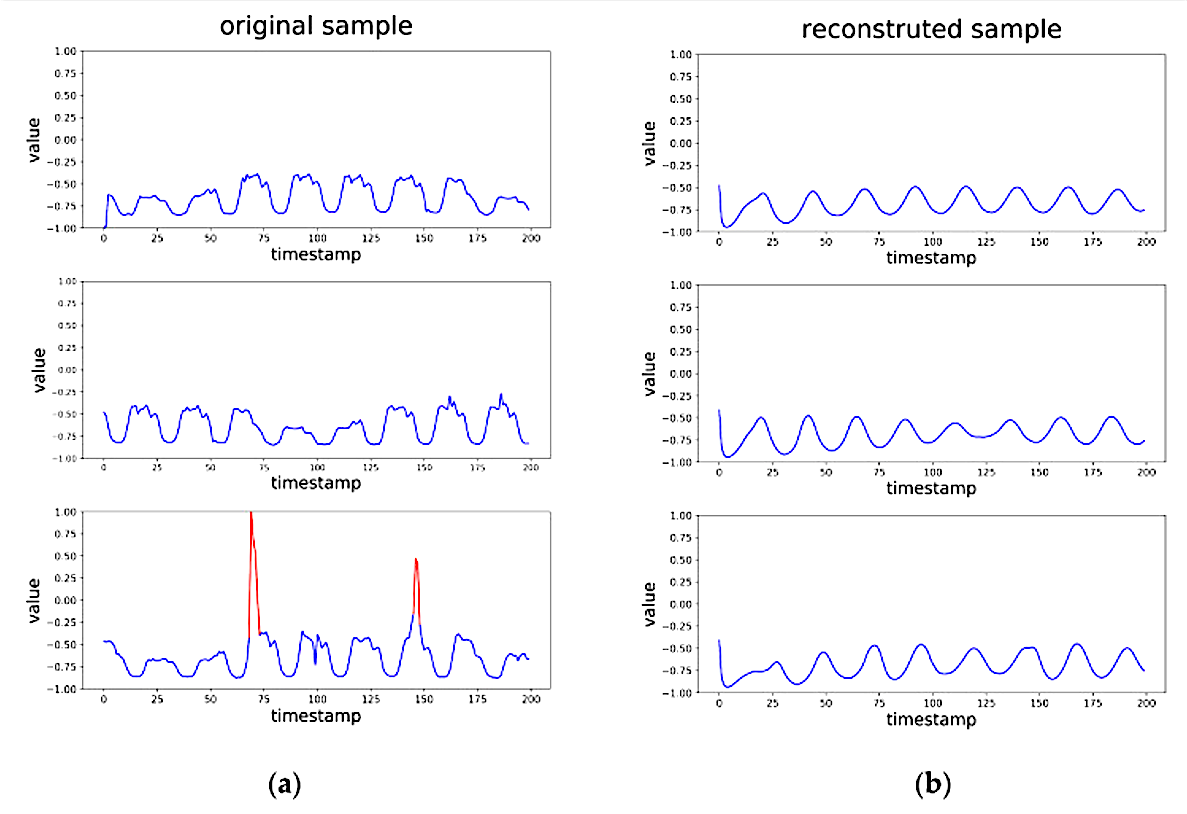}
    \caption{
    Toy example on autoencoders: After training, the autoencoder reconstructs inputs effectively when the test data shares the same distribution as the training data. However, in the presence of anomalies (red), the autoencoder fails to reconstruct them accurately, resulting in a spike in the reconstruction error at the anomaly's occurrence.}
    \label{method_ill}
\end{figure*}

In \cref{method_ill}, we illustrate the performance of an autoencoder (AE) using a toy example. The first two rows of column (a) show the input data used for training, and the corresponding reconstructed output is displayed in column (b). In the third row of column (a), we introduce artificial anomalies (colored red) into the input data. One notes that AE's reconstruction in column (b) shows significant deviations in regions corresponding to the anomalies. Computing the point-wise squared difference between the input and reconstructed outputs reveals distinct peaks at the locations of the anomalies.


\textbf{VAE-GAN.} The toy example provides a proof of concept that AE are good candidates for data with possible anomalies. In order to implement AE in the context of GW time series data, we propose an improved model based on variational autoencoders (VAE) \cite{kingma2013auto}. VAE are probabilistic models and their main advantage is anomaly detection even when the signals and the noise have the same mean (in which case the original AE might fail to perform well).

Similarly in our introduction of AE in \cref{ML}, VAE learns to reconstruct input data $x_i$ by maximizing the expected log-likelihood $\mathbb{E}_{\hat{x}\sim\text{VAE}(x)}[\log p(\hat{x}|\theta)]$ w.r.t. $\theta$. The process proceeds as follows: Given input data $x$, generate parameters $(\mu,\sigma) = \text{Encoder}(x)$. Then sample the latent variable $z$ from $q(z|x) \equiv\mathcal{N}(\mu,\sigma)$\footnote{Our prior for $z$ is $p(z)=\mathcal{N}(0,I_{d_Z\times d_Z})$. For numerical stability in gradient learning, one can apply re-parametrization trick to generate $z$ of the posterior distribution $q(z|x)$ by first sampling $\epsilon\sim\mathcal{N}(0,I_{d_Z\times d_Z})$, then computing $z=\mu+\sigma \odot \epsilon,$ where $\odot$ is an element-wise multiplication. Full details are to be found in \citep{kingma2013auto}.}. The decoding procedure is that given $z$ we learn $p(\hat{x}|z)$ by the same trick as before: we learn parameters $(\mu', \sigma') = \text{Decoder}(z)$ and sample $\hat{x}\sim\mathcal{N}(\mu', \sigma')\equiv p(\hat{x}|z)$. Both the Decoder and Encoder are neural networks parameterized by $\theta,$ (i.e. $p_\theta(\cdot|z)$) and $\phi$ (i.e. $q_\phi(\cdot|x)$) respectively. Recall that the training objective is maximizing the log-likelihood which can be shown to be equivalent to maximizing the following expression:
\begin{equation}\label{objective}
    \mathcal{L} = \sum_{i=1}^N \left( -D_{KL}(q_{\phi}(z|x^{(i)})||p(z)) + \frac{1}{L} \sum_{l=1}^L \log p_{\theta}(\hat{x}^{(i)}|z^{(i,l)}) \right)
    \end{equation}
where $D_{KL}$ is the KL divergence that measures distances between probability distributions, and $p(z)=\mathcal{N}(0,I)$ is our fixed prior for $z$.

\vspace{-0.5cm}
\subsection{Training Details}\label{training} 
The training/validation set contained $28800$ samples of noise-only data. The train/validation split was $90\% - 10\%$ respectively. A sample is a 4-second data at 4 kHz for each detector (we considered L1 and H1 detectors only as data from V1 detector was limited). Each sample then is a $4\times4000=16000$ entries concatenated into a one-dimensional array. With a sliding window of size $100$ and $50\%$ overlap, we generate $\lfloor\frac{16000-100}{50}\rfloor+1=319$ inputs each of size $100$ for each sample.

The data were whitened using a Fast Fourier Transform (FFT) to remove correlations in the noise \citep{cuoco2001line}. A band-pass filter (20-400 Hz) eliminated noise outside the interferometer's sensitivity range.

\begin{figure}
    \centering
    \includegraphics[width=1\linewidth]{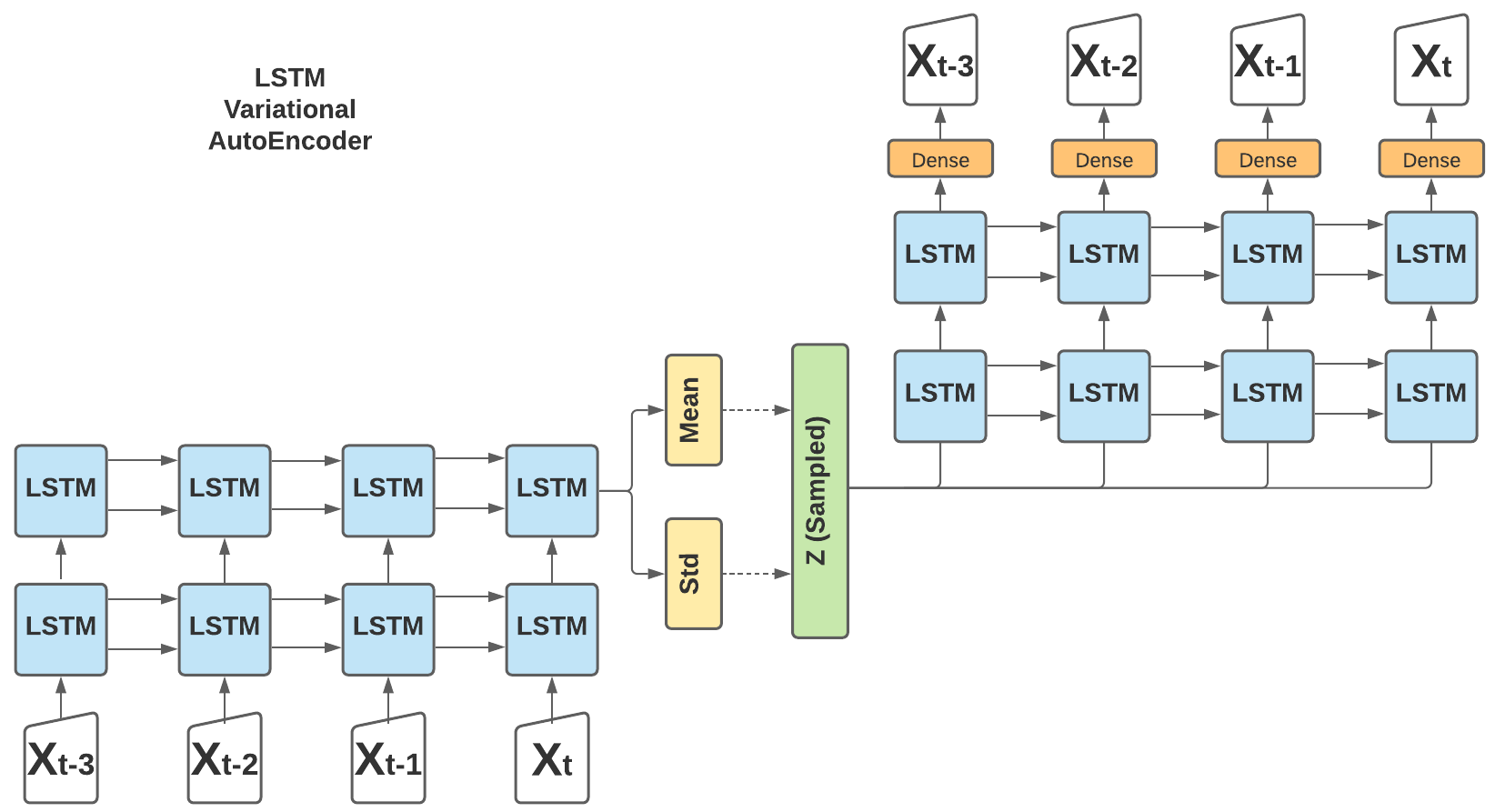}
    \caption{Illustration of our VAE model with LSTM units (not to scale). The encoder is the lower left block and the decoder is the upper right block. The objective \cref{objective} is maximized by back-propagating it through the network to find the optimal parameters $\theta,\phi$.}
    \label{vae-model}
\end{figure}

Shown in \cref{vae-model}, the encoder consists of two LSTM (Long short-term memory units) \citep{hochreiter1997long} layers of sizes 32 and 8 in that order. The final output of the LSTM units in the second layer is passed to two separate layers to produce $\mu$ and $\sigma$ respectively, each with dimension $d_z=8$. The decoder is two LSTM layers of sizes 8 and 32 respectively followed by layers to produce $\mu'$ and $\sigma'$ of the same dimension as the input.  The reason for choosing LSTM is their computational design to learn temporal dependencies. Other excellent choices may include temporal convolutional transformers. Learning was performed with Stochastic Gradient Ascent on the objective in \cref{objective}.

\begin{figure}
    \centering
    \includegraphics[width=1\linewidth]{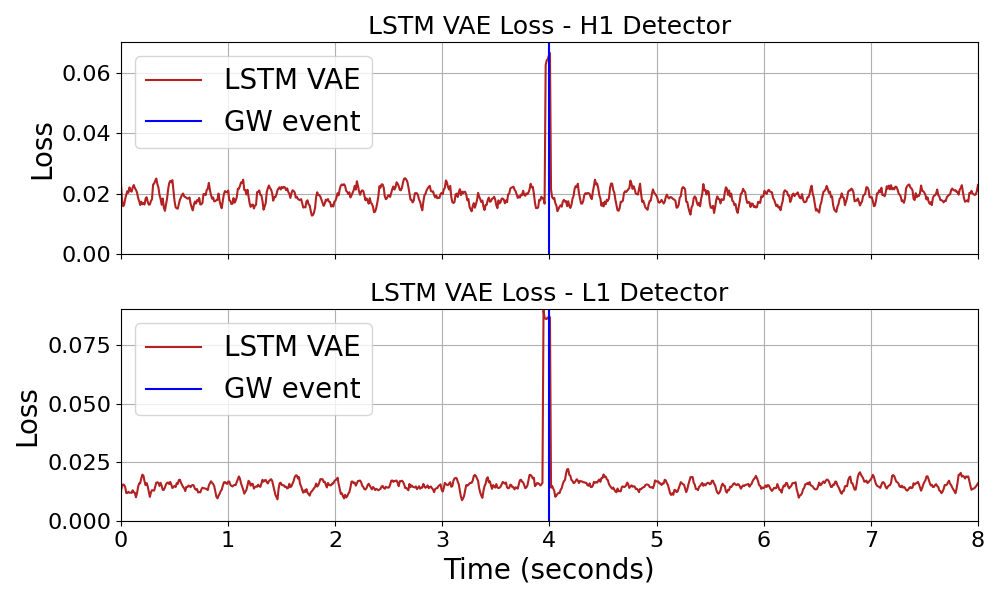}
    \caption{Performance on data from Event GW150914: quadratic loss on the stream of data. Notice how it peaks in the presence of the gravitational wave.}
    \label{perform}
\end{figure}
\vspace{-0.5cm}
\section{Results}
Testing data are a mixture of samples containing a signal and samples of pure noise. With the same procedure as in \cref{training}, we obtain the signal samples from events: GW \{150914, 170104, 170817, 190412, 190425, 190521, 190814, 200105, 200129, 200115\}.

In \cref{perform}, we show the performance of our VAE on Event GW150914 using data from both L1 and H1 detectors. Note that the squared distance between the input data and the VAE output (denoted by loss on the y-axis) is approximately stationary and close to $0$ in the existence of regular noise. However, once the GW passes through the detectors, a spike in the loss manifests, indicating the presence of an anomaly.

\begin{figure}[H]
    \centering
    \includegraphics[width=0.85\linewidth]{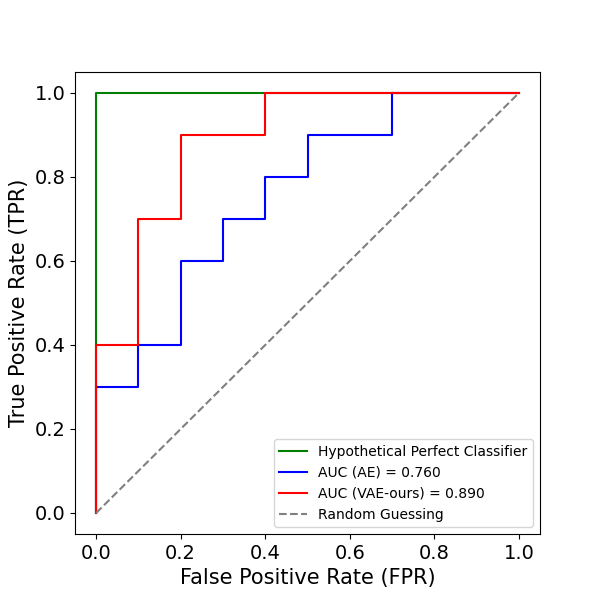}
    \caption{ROC curve analysis with 10 positive (GW events) and 10 negative (noise) samples. Note how the VAE's performance exceeds that of a vanilla AE.}
    \label{roc}
\end{figure}

For input data, we declare them anomalous if the anomaly score is above a certain threshold $\alpha$; the anomaly score is simply the distance between the input of the VAE and its output. For each threshold, we compute the true positive rate (TPR, or recall) and the false positive rate (FPR) for our testing data. By varying the threshold we obtain the Receiver Operator Characteristic (ROC) curve. The thresholds are found using the python command $\texttt{sklearn.metrics.roc\_curve}$. It is well-known in statistics that the area under the ROC curve (AUC) is a measure of the goodness of a classifier, where a higher value (e.g. $>0.7$) indicates a better performance. \Cref{roc} illustrates the ROC curve on the 10 GW event data and 10 noise-only data. The AUC was found to be $\pmb{0.89}$. We also report the F1 score to be $0.857$, which is the harmonic mean of the precision and recall (the higher the better).

\vspace{-1.3cm}
\section{Conclusion}
\vspace{-0.3cm}
This study introduced an unsupervised learning approach to detect anomalies in gravitational wave time-series data, utilizing variational autoencoders (VAEs). The method demonstrated robust performance in identifying anomalies, such as GW150914, by detecting spikes in reconstruction error. The method's performance was assessed using ROC analysis with an AUC of 0.89. To our knowledge, this is the first work to use VAE in the context of astrophysics applications. Due to time constraints, we were not able to perform further studies (e.g. comparison to \citep{raikman2023gwak}). We will leave that to future work.
The proposed method could be used in different scenarios due to its unsupervised nature to detect possibly new physics in certain datasets with no given templates.

We gratefully acknowledge Gunther Roland for his valuable feedback and insightful comments on this work.

\bibliography{ref}

\end{document}